\documentclass[superscriptaddress,
reprint,longbibliography,
preprintnumbers,nofootinbib,
amsmath,amssymb,aps,prl]{revtex4-1}

\pdfoutput=1

\usepackage[utf8]{inputenc} 
\usepackage[T1]{fontenc} 
\usepackage{microtype}
\usepackage{mathrsfs}
\usepackage{amsmath}

\usepackage{graphicx}

\usepackage[dvipsnames]{xcolor}

\usepackage{dcolumn}
\usepackage{bm}

\usepackage[colorlinks = true,
            linkcolor = blue,
            urlcolor  = blue,
            citecolor =black,
            anchorcolor = blue]{hyperref}
\usepackage{verbatim}
\usepackage{color,ulem}
\usepackage[english]{babel}

\usepackage[scr]{rsfso}

\newcommand{\cP}{\mathcal{P}}

\usepackage{blindtext}

\begin{document}

\preprint{IFT-UAM/CSIC-20-8}

\title{\Large Homogeneous Holographic Viscoelastic Models \& Quasicrystals}

\author{Matteo Baggioli}%
 \email{matteo.baggioli@uam.es}
\affiliation{Instituto de Fisica Teorica UAM/CSIC, c/Nicolas Cabrera 13-15,
Universidad Autonoma de Madrid, Cantoblanco, 28049 Madrid, Spain.}%


\begin{abstract}
We show that the field theories dual to the homogeneous holographic models with spontaneously broken translations display several distinctive properties of \textit{quasicrystals}, aperiodic crystals with long-range order (e.g. incommensurate charge density waves). This interpretation suggests that the longitudinal diffusive mode, observed in the spectrum of excitations of these systems, is the diffusive Goldstone boson typical of \textit{quasicrystals} -- the \textit{phason}. Moreover, following this idea, and using the effective field theory formalism for Goldstone modes in dissipative systems, we are able to derive the universal phase relaxation found in these holographic models in presence of a small source of explicit breaking (\textit{pseudo-spontaneous} regime).
\end{abstract}

\maketitle

\section{Introduction}

In the current literature, there are several types of holographic models which break translational invariance while retaining a homogeneous geometry. We can summarize them into three classes: massive gravity \cite{Vegh:2013sk, Andrade:2013gsa, Baggioli:2014roa, Alberte:2015isw}, Q-lattices \cite{Donos:2013eha, Amoretti:2017frz} and helical lattices \cite{Andrade:2017cnc, Andrade:2018gqk}. All of them rely on some sort of global symmetry, usually a $U(1)$ shift or a more exotic combination of translations and rotations \cite{Iizuka:2012pn}. These setups appear interesting because of their potential toy model role to understand strange metals and strongly correlated solids \cite{Amoretti:2017axe, Amoretti:2018tzw, Davison:2013txa, Delacretaz:2016ivq}.\\
Regardless of the intense activity on every front, most of the fundamental questions are still open. As a concrete example, the correct hydrodynamic description \cite{Armas:2019sbe} of those models has been confirmed only recently \cite{Ammon:2020xyv}, in contrast to the previously accepted framework \cite{Delacretaz:2017zxd} which was clearly in tension with the holographic results \cite{Ammon:2019apj,Donos:2019hpp}.\\[0.1cm]
An even more fundamental question is left. \color{black}\textit{Which phase of matter are these holographic models describing} ?\color{black}\\[0.1cm]
The confusion in the literature is manifest: Wigner crystals \cite{Amoretti:2018tzw}, metallic density waves \cite{Amoretti:2018tzw}, viscoelastic solids \cite{Andrade:2019zey,Baggioli:2019mck}, charge density waves \cite{Amoretti:2017frz}, amorphous solids \cite{Ammon:2019wci}, strange insulators \cite{Andrade:2018gqk}, scale invariant solids \cite{Baggioli:2019elg}, spontaneous helices \cite{Andrade:2017cnc}, homogeneous lattices \cite{Andrade:2015iyf}, strange metals with slowly fluctuating translational order \cite{amoretti2018holographic}, states with dynamical defects \cite{Grozdanov:2018ewh}, holographic lattices \cite{Donos:2017ihe}. Probably, the only indisputable statement is that these phases break spontaneously translations, as suggested in the title of \cite{Donos:2018kkm}.\\

\color{black}Most of this confusion arises because of the presence of the aforementioned global symmetries within the gravitational bulk dynamics. From this perspective, global symmetries come as unfamiliar and odd for several reasons. First, global symmetries are believed to be absent in a consistent and UV complete theory of quantum gravity \cite{Harlow:2018jwu}. Nevertheless, here we are taking a fully low energy bottom-up approach, in which this point should not be considered as an impediment. The global symmetries in these models could just be accidental -- emergent symmetry of the IR dynamics. The problem of embedding these setups into UV complete constructions goes far beyond the scope of this paper and moreover it is not clear how relevant it is in order to apply these frameworks to condensed matter systems. Second, and most importantly, the holographic dictionary is based on the slogan that ''\textit{global symmetries at the boundary correspond to gauge symmetries in the bulk}'' \cite{Baggioli:2019rrs}. What global symmetries in the bulk correspond to at the boundary? To the best of our knowledge, no convincing solution to this problem has been presented so far. Nevertheless, there are explicit holographic constructions \cite{Amado:2013xya} which analyze this problem and surprisingly found that the features of the boundary Goldstone dynamics is independent of whether the symmetry is gauged or global in the bulk. Finally, connecting to the previous point, the work of \cite{Esposito:2017qpj} noticed that keeping the internal symmetries global in the bulk does not correspond to have the standard symmetries of solids (as envisaged in the EFT framework of \cite{Nicolis:2015sra}) at the boundary, for which on the contrary those internal symmetries should be gauged in the bulk as done in \cite{Esposito:2017qpj}.\\

Once the previous points are accepted, one possibility is to think about these global bulk symmetries as spurious artifacts of these simplified toy models and just discard all the features coming from their presence (existence of extra longitudinal diffusive mode, phase relaxation $\Omega$, etc \dots). Our approach, is somehow opposite, and it is more into the direction of trying to give them physical meaning from a condensed matter point of view.\color{black}\\

In this work, we propose that a viable interpretation for these holographic models is that of \textit{quasicrystals}. To motivate this picture more precisely, let us consider the following features shared by all the holographic models and compatible with our interpretation.
\begin{enumerate}
\setlength\itemsep{0.05em}
    \item Spacetime translations are not broken to a discrete subgroup; there is no unit cell. The systems are not periodic.
    \item The systems are rotationally invariant and can be also scale invariant. As we will see, most \textit{quasicrystals} are invariant under \color{black}discrete \color{black} rotational groups (e.g. $5$-fold symmetry \cite{Caspar14271}) and some of them (e.g. Penrose tilings \cite{penrose1974role}) are self-similar \color{black}-- invariant under discrete scale symmetries. Penrose tiling have been indeed already considered as discrete models of conformal geometries \cite{PhysRevX.10.011009}.\color{black}
    \item The longitudinal spectrum contains a diffusive Goldstone mode, which does not generate from the breaking of spacetime translations \cite{Donos:2019txg}. Moreover, its diffusion constant goes to zero with the temperature, confirming its dissipative nature. This is the \textit{phason} of the \textit{quasicrystal}\footnote{This interpretation was already contemplated in the context of incommensurate charge density waves in \cite{Donos:2019txg}.}.
    \item There are no commensurability effects in these systems \cite{Andrade:2015iyf}, in contrast to the inhomogeneous (and periodic) holographic lattices \cite{Andrade:2017leb,Andrade2018}. Incommensurate density waves are indeed one of the most common examples of \textit{quasicrystals} \cite{scott1987incommensurate}.
    \item Most of these holographic systems are meta-stable -- not thermodynamically favoured states. \color{black} More precisely, at the present time, all the homogeneous holographic models which are dynamically stable are not thermodinamically stable \cite{Ammon:2020xyv}. It is tempting to think about the existence of a physical reason behind this fact. Simultaneously, the stability of \textit{quasicrystals} is still a controversial topic and most of them are believed to be meta-stable phases of matter \cite{doi:10.1080/14786430500419411}. This connection might give a physical meaning to the meta-stability of the holographic models as well.\color{black}
\end{enumerate}

Using this new understanding, (I) we are able to explain the physical nature of the longitudinal diffusive mode observed in the spectrum \cite{Ammon:2019apj,Baggioli:2019abx,Donos:2019txg,Amoretti:2018tzw} as the common \textit{phason} mode of \textit{quasicrystals}; (II) we are able to prove \color{black}in the decoupling limit \color{black} the universal relation for the phase relaxation proposed in \cite{Amoretti:2018tzw}, and confirmed in \cite{Donos:2019txg,Baggioli:2019abx,Ammon:2019wci,Amoretti:2019kuf,Amoretti:2019cef}, using effective field theory methods for diffusive goldstone bosons in dissipative environments \cite{Minami:2018oxl,Hidaka:2019irz,Landry:2019iel}. \color{black}Moreover, we provide strong evidence for its validity beyond the decoupling limit. \color{black}
\section{Quasicrystals and phasons, a brief primer}
Before 1984 \cite{PhysRevLett.53.1951}, everyone was assuming that all solids are crystals composed of a periodic arrangement of identical unit cells. In other words, there was no distinction between two fundamental concepts: long-range order and periodicity. The discovery of \cite{PhysRevLett.53.1951} introduced the concept of \textit{quasicrystal} \cite{PhysRevLett.53.2477}, as a totally new kind of long-range order (for more details see \cite{jaric1988introduction,janssen2007aperiodic,janssen1988aperiodic}).\\

\begin{figure}[h!]
    \centering
    \includegraphics[height=1.5cm]{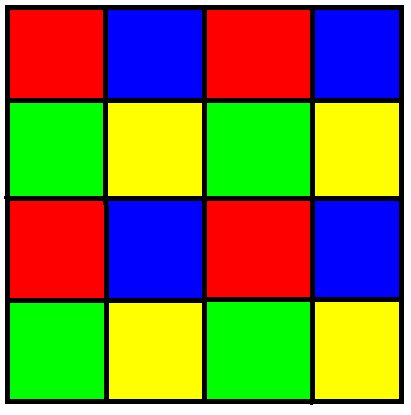} \quad 
      \includegraphics[height=1.5cm]{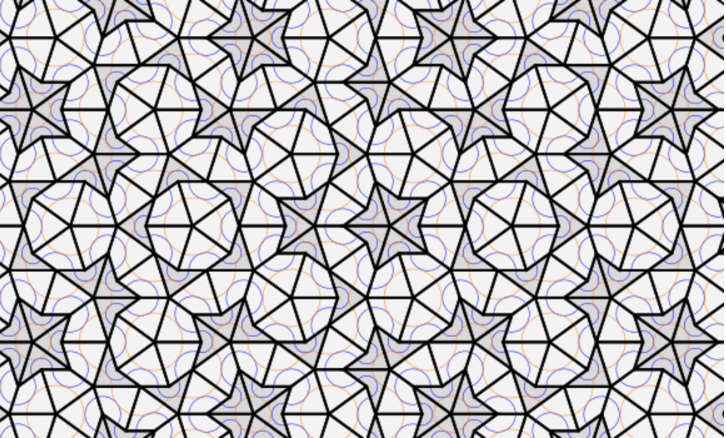}
      \quad 
      \includegraphics[height=1.5cm]{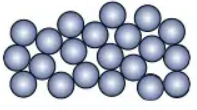}
    \caption{A periodic tiling (crystal) versus an aperiodic Penrose tiling (\textit{quasicrystals}) versus a disordered tiling (amorphous crystal).}
    \label{fig1}
\end{figure}

The differences between crystals and \textit{quasicrystals} can be visually understood thinking of tilings. It is simple to imagine a kitchen floor with square tiles; it is harder to think about non-periodic tessellations of the plane. The distinction is manifestly visible in fig.\ref{fig1}: aperiodic tilings lack completely any form of discrete translational invariance and very often enjoy \color{black} discrete \color{black} rotational invariance. Even more curiously, some of them -- Penrose tilings \cite{Penrose1974TheRO} -- are self-similar; the same patterns occur at larger and larger scales. In our language, those non-periodic structures are scale invariant, exactly as the holographic models which we will discuss\footnote{\color{black}As already emphasized, the scale invariance of Penrose tilings is discrete and not continuous as in the holographic constructions. \color{black}}.
\\
The fundamental point of the experimental discovery of quasicrystals \cite{PhysRevLett.53.1951} was the lack of (discrete) translational symmetry in their Bragg  diffraction patterns. Nevertheless, the diagrams displayed sharp Bragg peaks, signaling the presence of long-range order. In this regard, \textit{quasicrystals} are definitely different with respect to glasses and amorphous systems, where long-range order is absent. Notice, that quasicrystals, despite breaking translational invariance, might possess discrete rotational symmetries (e.g. $5-$ fold symmetry), \color{black}which furthermore may be incompatible with periodicity.\color{black}\\

Interestingly enough, the differences between crystals and \textit{quasicrystals} are not limited to their diffraction patterns but are also evident in their dynamics, even in the hydrodynamic regime, intended as small frequency and momentum with respect to the temperature. More precisely,
\textit{quasicrystals} display new hydrodynamic modes (which can be thought as additional Goldstone modes) known as \textit{phasons} \cite{jaric1988introduction,janssen2007aperiodic,janssen1988aperiodic}, whose dispersion relation is diffusive:
\begin{equation}
    \omega\,=\,-\,i\,D_{phason}\,k^2\,+\,\dots
\end{equation}
These modes \textit{do not} come from the spontaneous breaking of translational invariance; they are fundamentally different from \textit{phonons}, to which they are nevertheless coupled. \textit{Phasons} and their diffusive nature can be experimentally observed \cite{PhysRevLett.108.218301,PhysRevLett.57.1440,doi:10.1002/ijch.201100131}. Ultimately, \textit{phasons} are related to specific rearrangements of the atomic structure and they have an activated nature, namely their diffusion constant vanishes at zero temperature.\\

\begin{figure}
    \centering
    \includegraphics[width=0.45 \linewidth]{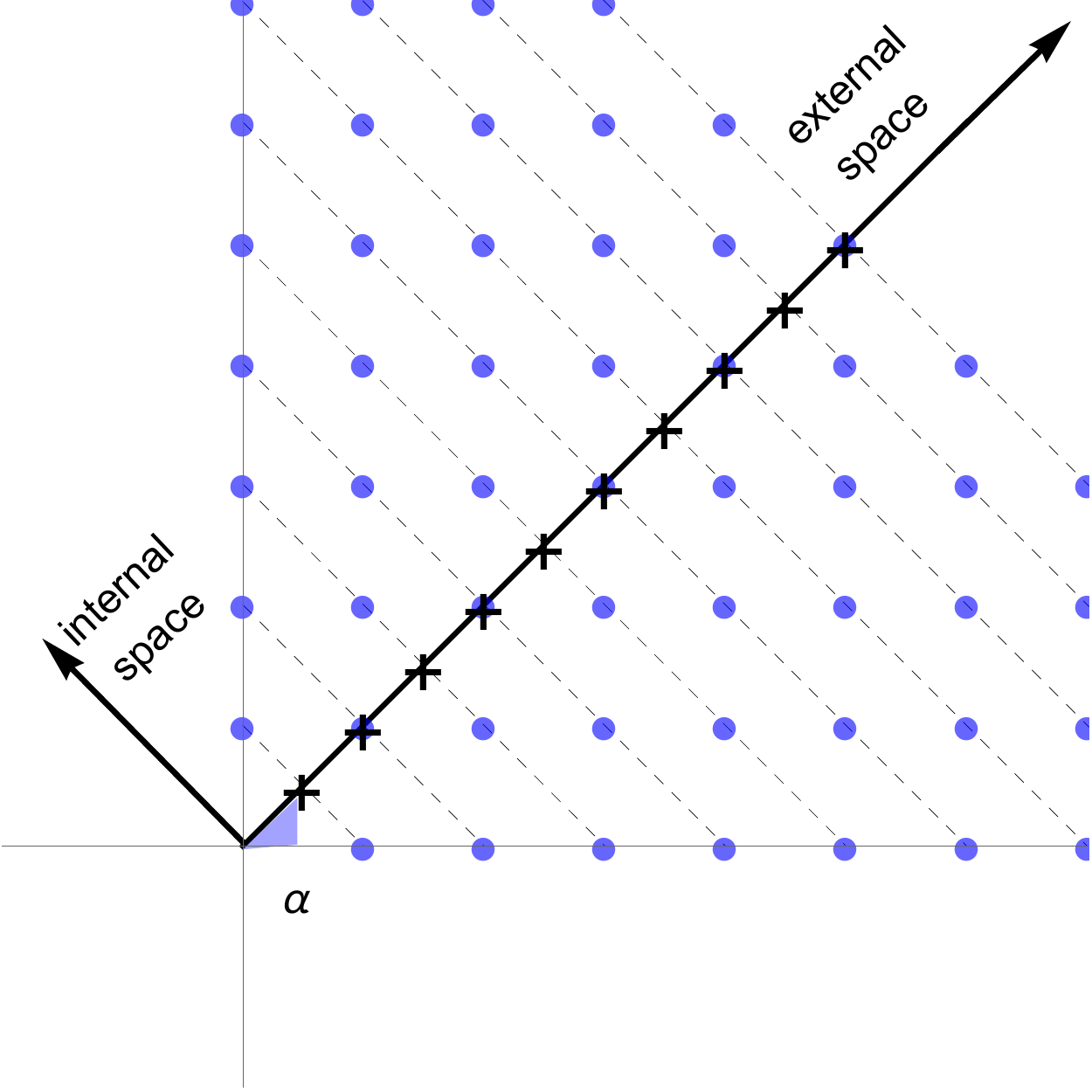}
    \quad 
    \includegraphics[width=0.45 \linewidth]{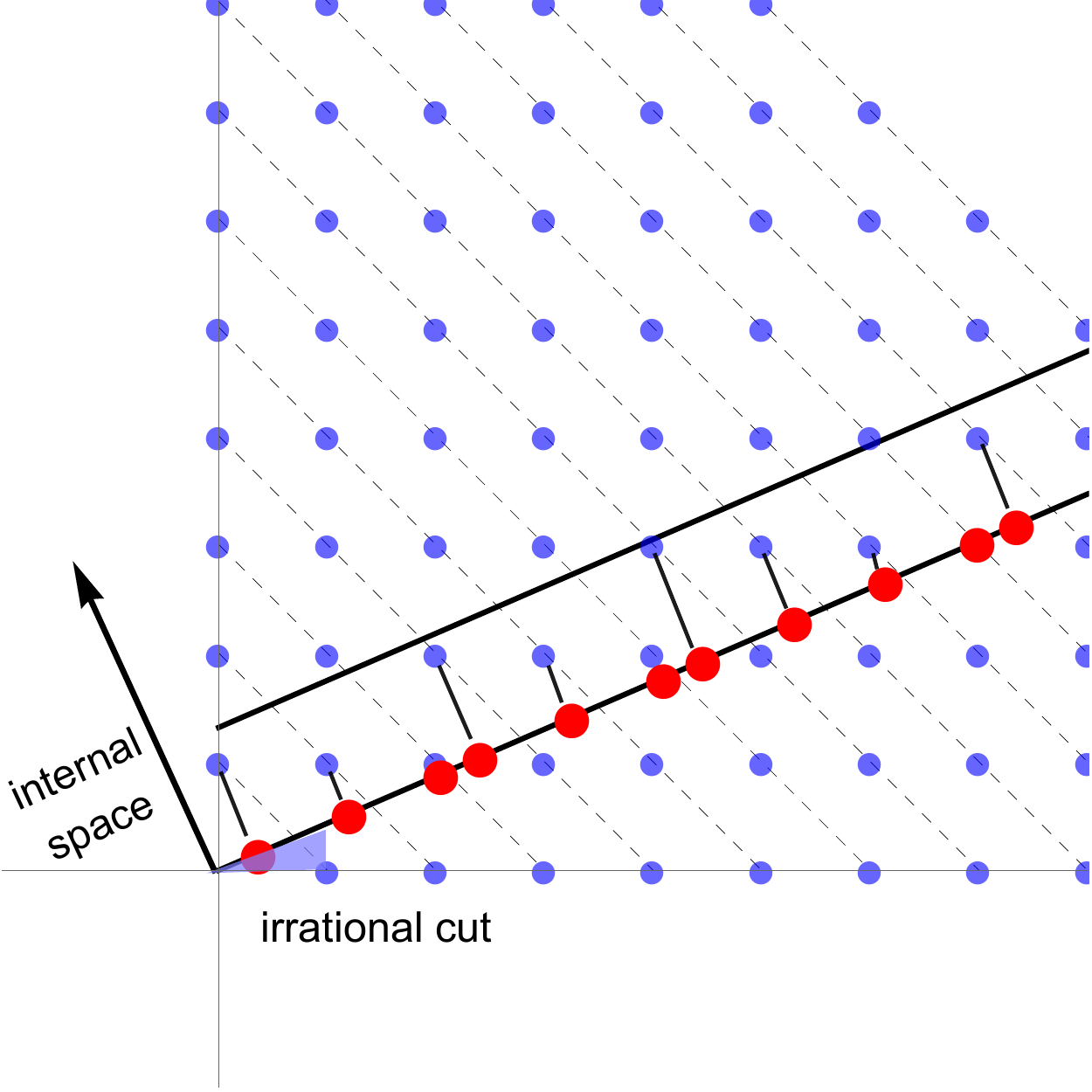}
    \caption{The superspace representation of a periodic lattice and an aperiodic lattice in $1D$. The higher dimensional lattice (in this case $2D$) is always periodic. In the first case the cut is rational, while in the second one is irrational.}
    \label{nonp}
\end{figure}

In order to understand precisely what a \textit{phason} is, it is convenient to adopt the \textit{superspace} picture introduced by Bohr \cite{bohr1925} \footnote{\color{black}Not to be confused with the superspace formalism of supersymmetric theories nor the mini superspace of quantum gravity frameworks.\color{black}}. More specifically, aperiodic crystals can be obtained from the section of a periodic crystal of higher dimension cut at an irrational angle. See figure \ref{nonp} for a $2D \rightarrow 1 D$ representation. Phonon displacements modify the position of the lattice points in ''real'' space; on the contrary, phasons change the position with respect to the quasi-crystal structure and the (irrational) cut that determines it. In other terms, phonon modes are excitations of the real (also called parallel or external) space whereas phasons are excitations of the perpendicular (or better internal) space. This extra-dimension picture will re-appear in the homogeneous holographic setups.\\

In order to gain some intuition, it is helpful to think about the simplest incommensurate structure given by two lattices superimposed with a modulation\footnote{Typical examples are incommensurate charge density waves, experimentally observed in Cuprates \cite{Torchinsky2013}. This is exactly the physical picture behind the hydrodynamic theory of \cite{Delacretaz:2017zxd} and the holographic discussions in \cite{Donos:2019txg}.}. The free energy of the system is invariant under a phase shift of the modulation; that is the origin of the \textit{phason} \cite{van2007incommensurate}.
\section{The structure of the holographic models}
\begin{figure}
    \centering
    \includegraphics[width=0.6 \linewidth]{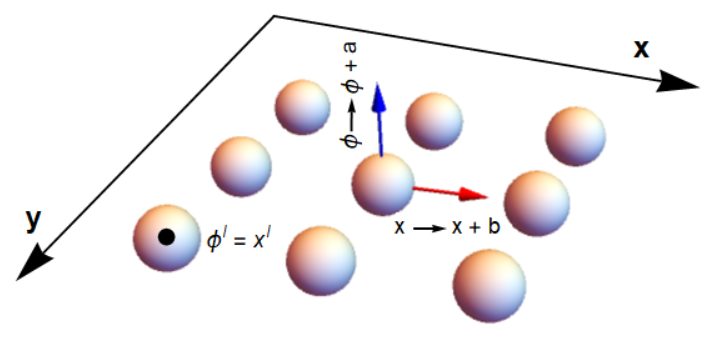}
    \caption{A \textit{superspace} description of our holographic setup. The shift in the internal space of the scalar fields $\phi^I$ can be thought as a standard translation in a spacetime with higher number of dimensions -- the \textit{superspace}.}
    \label{cartoon}
\end{figure}

Let us be more precise about the symmetry structure of the holographic models which we are after. For simplicity, we will focus on two specific classes, defined by the following bulk scalar fields:
\begin{equation}
    \textit{a)}\,\,\,\,\phi^I=\alpha\, x^I\,,\quad \quad \textit{b)}\,\,\,\,\phi\,=\,e^{i k \cdot x}\,\varphi
\end{equation}
which break the invariance under spacetime translations of the dual field theory. Depending on the boundary conditions \cite{Alberte:2017oqx,Amoretti:2017frz,Armas:2019sbe}, the breaking can be explicit \cite{Andrade:2013gsa}, spontaneous \cite{Alberte:2017oqx,Andrade:2019zey,Baggioli:2019elg,Baggioli:2019mck} or pseudo-spontaneous \cite{Alberte:2017cch,Andrade:2017cnc,Ammon:2019wci,Baggioli:2019abx}.
In the first class \cite{Baggioli:2014roa,Alberte:2015isw}, a finite value of $\alpha$ breaks two different set of symmetries: spacetime translations (and rotations) $x^I \rightarrow x^I+ b^I$ and internal global translations $\phi^I \rightarrow \phi^I+a^I$. More precisely, both symmetries are broken to the diagonal subgroup $a^I=-b^I$, exactly in the same way as in the EFTs of \cite{Nicolis:2015sra}. In the second class \cite{Donos:2013eha} the SSB is induced by the formation of a finite vev $\langle \varphi \rangle$ and one can engineer two different symmetry breaking patterns, which are now separable \cite{Donos:2019txg,Amoretti:2018tzw}. When $k$ is zero, only the global $U(1)$ symmetry is broken; while for $k \neq 0$ spacetime translations and the global $U(1)$ are broken again to the diagonal group.\\
The important thing to notice is that in both models translations is not the only broken symmetry. There is always a broken global symmetry in the dual field theory; and most importantly new Goldstone modes associated exclusively to that \cite{Donos:2019txg}. This is related to the existence of an additional \textit{internal} direction, ''transversal'' to the real spacetime, which is shown in fig.\ref{cartoon} for model $a)$. Because of the shift invariance of the model, shifts in the internal space of the scalars $\phi^I$ do not cost energy. This picture is totally analogous to the superspace description of quasicrystals in the previous section. The dynamics in that ''extra-dimension'' is exactly what gives rise to the extra Goldstone mode -- the \textit{phason}.
Interestingly, also \textit{quasycristals} are left invariant by a combination of phason and phonon displacements at which lattice points are not moving. This combination is referred to as the \textit{characteristic displacement} and it corresponds exactly to the lattice vector of the \textit{hypercrystal} living in the \textit{superspace}. In summary, it is tempting to associate the scalar shifts to the \textit{phason} displacements and the diagonal preserved group to the \textit{characteristic} one.

\section{The longitudinal diffusive mode as a phason}
In all the homogeneous holographic models, the longitudinal spectrum contains a diffusive mode with dispersion relation \cite{Armas:2019sbe}:
\begin{equation}
    \omega\,=\,-\,i\,D_\parallel\,k^2\,+\,\dots\,,\quad D_\parallel\,=\,\xi\,
  \frac{\left(B+G-\cP\right)\,\chi_{\pi\pi}}{s'\,T^2\,v_\parallel^2}\label{cc}
\end{equation}
where $s$ is the entropy density, $s'\equiv ds/dT$, $B$ the bulk modulus, $G$ the shear modulus, $\chi_{\pi\pi}$ the momentum susceptibility, $\mathcal{P}$ the configuration pressure and $v_\parallel$ the speed of longitudinal phonons. Eq.\eqref{cc} has been recently tested explicitly in a large class of models \cite{Ammon:2020xyv}. \color{black}The agreement with hydrodynamics confirms that this mode is connected to mass motion and the thermally activated rearrangement of the atomic structure \cite{PhysRevB.13.500}.\footnote{\color{black}Nevertheless, it is not related to the motion of vacancies, dislocations or defects which are clearly absent in the holographic picture.\color{black}} \color{black} Most importantly, the diffusion constant crucially depends on the dissipative parameter $\xi$, which is derived from the Green function of the scalar operators $\Phi$ as:
\begin{equation}
    \xi
    = -\,\lim_{\omega\to0}\lim_{k\to0}
    \omega\,\mathrm{Im}\,G^R_{\Phi\Phi}
\end{equation}
This clarifies that such diffusive mode can exist only in dissipative systems at finite temperature; this is the reason why it was never discussed in the EFT framework of \cite{Nicolis:2015sra}. \footnote{In \cite{Nitta:2017mgk,Musso:2019kii,Musso:2018wbv,Gudnason:2018bqb}, the possibility of having additional and independent Goldstone modes, because of the breaking of the global $U(1)$ symmetry, was already proposed. Nevertheless, because of the absence of dissipation, the diffusive nature was not revealed. }\\
From a symmetry perspective, it was shown in \cite{Donos:2019txg,Amoretti:2018tzw} that such mode comes from the spontaneous breaking of the internal global symmetry of these models. Moreover, this diffusive mode can be obtained in the decoupling limit \cite{Ammon:2019apj,Baggioli:2019rrs,Baggioli:2019lsz,Amoretti:2018tzw}, in which the momentum operator plays no role. A possible simple way to understand the nature of this diffusive mode is by thinking that a ''real'' solid should enjoy $\text{ISO}(d)$ global symmetry \cite{Nicolis:2015sra}. According to the holographic dictionary, this means that $\text{ISO}(d)$ should be gauged in the bulk, as done in \cite{Esposito:2017qpj}. If one proceeded with the gauging, the diffusive modes would be simply associated to the corresponding conserved currents.
\section{The universal phase relaxation from effective field theory}
In presence of a small explicit breaking $\langle EXB \rangle$, standard propagating Goldstone bosons acquire a mass gap (or in condensed matter language, a \textit{pinning frequency}) $\omega_0$, which obeys the famous Gell-Mann-Oakes-Renner (GMOR) relation \cite{PhysRev.175.2195}, namely:
\begin{equation}
    \omega_0^2\,\sim\,\langle EXB\rangle \,\langle SSB \rangle 
\end{equation}
This mechanism, well-know in particle physics because of the Pions, it is reproduced in a standard way in the holographic models \cite{Ammon:2019wci,Andrade:2018gqk,Amoretti:2018tzw,Amoretti:2016bxs}, even for spacetime symmetries. Additionally, the longitudinal diffusive mode, acquires a finite relaxation rate $\Omega$ \cite{Donos:2019txg}, appearing in its dispersion relation and entering its Josephson relation \cite{Delacretaz:2017zxd}.
This term is fundamentally different with respect to that induced by elastic defects, such as dislocations, and relevant for the discussions of \cite{Delacretaz:2016ivq,Delacretaz:2017zxd}, because it depends crucially on the explicit breaking scale.
More importantly, \color{black} in the context of holography, \color{black} it has been conjectured \cite{Amoretti:2018tzw}, and verified numerically \cite{Amoretti:2018tzw,Andrade:2018gqk,Amoretti:2019cef,Amoretti:2019kuf,Donos:2019txg,Donos:2019hpp,Baggioli:2019abx,Ammon:2019wci}, that such relaxation rate obeys the universal relation:
\begin{equation}
    \Omega\,=\,\omega_0^2\,\xi\,\chi_{\pi\pi} \label{toprove}
\end{equation}
As a matter of fact, no fundamental explanation for this expression appeared so far.\\
Here, we use the EFT methods presented in \cite{Minami:2018oxl,Hidaka:2019irz} to show that Eq.\eqref{toprove} is a standard result for diffusive Goldstone modes in presence of an explicit breaking source.\\
Let us start by using the notations of \cite{Hidaka:2019irz} and by defining the most general structure for the Goldstone modes (matricial) Green function:
\begin{equation}\label{exp}
    \left[G_\Phi^{-1}(k)\right]^{ab}\,=\,C^{ab}\,-\,i\,C^{ab;\mu}\,k_\mu\,+\,C^{ab;\mu\nu}\,k_\mu k_\nu\,+\,\dots
\end{equation}
where $k^\mu=(\omega,\vec{k})$ is the four-momentum.\\
In absence of explicit breaking $C^{ab}=0$, while in absence of dissipation (e.g. $T=0$ field theories) $C^{ab;0}$ reduces to the Watanabe-Brauner matrix $\rho^{ab}$ in the EFT description of \cite{Watanabe:2012hr,Watanabe:2014fva}. In any case, a non trivial $C^{ab;\mu}$ matrix (with rank different from zero) is needed to obtain type B Goldstone modes \cite{Watanabe:2019xul}. The novelty of \cite{Minami:2018oxl,Hidaka:2019irz} is noticing that, in dissipative environments, the type B Goldstone can be not only propagating ($\omega= A k^2$) but also diffusive ($\omega=-i D k^2$).\footnote{Similar results have been recently achieved in \cite{Landry:2019iel} using Coset techniques.}\\
Continuing, in absence of EXB, the Green function of the scalar operator at zero momentum reads \cite{Amoretti:2018tzw}:
\begin{equation}
    G_{\Phi\Phi}(\omega)\,=\,\frac{1}{\chi_{\pi\pi}\,\omega^2}\,-\,\xi\,\frac{i}{\omega}\,+\,\dots\label{ee}
\end{equation}
and has been checked numerically in various holographic models \cite{Ammon:2019apj,Ammon:2019wci,Amoretti:2018tzw,Amoretti:2019cef,Amoretti:2019kuf}. \color{black}Working in the decoupling limit (where the first term in \eqref{ee} is absent) \cite{Ammon:2019apj,Donos:2019txg,Amoretti:2018tzw} \color{black}, and using the diffusive dispersion relation $\omega=-iD k^2$, we can fix uniquely the coefficients appearing in eq.\eqref{exp} as:
\begin{equation}
    C^{;0}\,=\,1/\xi,\quad C^{;xx}\,=\,D/\xi\,,\label{param}
\end{equation}
where we have omitted the indices $a,b$, since for simplicity we are considering a single Goldstone. \color{black} When the $\Phi$ operator couples to all the other excitations, as in \cite{Donos:2019hpp}, the dynamics is modified, but only at higher order in the frequency.\footnote{\color{black}Notice also how eq.\eqref{ee}, which is valid only in the coupled system, cannot be reproduced by a single field Green function of the type \eqref{exp}. Moreover, taking the decoupled Green function as in \eqref{ee} would give rise to an unphysical pole $\omega=-i \chi/\xi$.\color{black}} \color{black} Therefore, from now on we will consistently truncate our expressions to order $\mathcal{O}(\omega)$.\\ It is now straightforward to add a pinning frequency $\omega_0$ to the Green function in \eqref{ee}:
\begin{equation}
    C\,=\,-\,\omega_0^2\,\chi_{\pi\pi}\label{mass}
\end{equation}
which appears as a result of a small explicit breaking (GMOR). Now, from the Green function in eq.\eqref{exp} and using the parameters in \eqref{param}$+$\eqref{mass}, we can obtain the dispersion relation for the (now) pseudo-diffusive mode, in presence of a small EXB, as:
\begin{equation}
    \omega\,=\,-\,\xi\,\omega_0^2\,\chi_{\pi\pi}\,+\,\dots
\end{equation}
Finally, the latter implies that the phase relaxation rate is given exactly by eq.\eqref{toprove} -- the universal value discussed in \cite{Amoretti:2018tzw}. In summary, this universal expression, which follows directly from \cite{Hidaka:2019irz}, is not surprising and it is just a consequence of the Goldstone nature of the diffusive mode.\\
Notice, that also the other terms entering in eq.\eqref{exp} would get corrected by the explicit breaking scale, but only at order $\langle EXB \rangle^2$ (e.g. the momentum relaxation rate $\Gamma$). At leading order, these corrections can therefore be neglected with respect to $\omega_0^2 \sim \langle EXB \rangle$.
In summary, our results are valid only at order $\mathcal{O}(\omega)$ and $\mathcal{O}(\langle EXB\rangle)$; fortunately, this is sufficient to identify the relaxation rate $\Omega$.
Let us stress that, because of the lock-in between internal shifts and spacetime translations, and the necessity of having explicit breaking (to have a finite $\Omega$), the decoupling approximation does not hold generally. This means, for example, that we do not expect the dispersion relation of the pseudo-diffusive \textit{phason} to be simply $\omega=-i \Omega-i D k^2$,
but rather to show a complicated interplay with the other modes as already discussed in \cite{Baggioli:2019abx,Donos:2019hpp}. As a concrete example, it is known \cite{Ammon:2019wci,Amoretti:2018tzw,Amoretti:2019cef} that the pseudo-diffusive mode in the transverse sector collides with the Drude pole $\omega=-i \,\Gamma$, creating the expected gapped \textit{pseudo-phonons}. Nonetheless, the relation for the relaxation rate $\Omega$, which can be extracted numerically from the dynamics of the \textit{pseudo-phonons}\footnote{More precisely, at zero momentum $k=0$ and small explicit breaking, this dynamics is described by the zeros of the following expression:
\begin{equation}
    \left(\Omega\,-\,i\,\omega\right)\,\left(\Gamma\,-\,i\,\omega\right)\,+\,\omega_0^2\,=\,0\,,
\end{equation}
where $\Gamma$ is the momentum relaxation rate.}, is going to be, at leading order ($\langle EXB \rangle \ll 1$), totally insensitive to those couplings and in agreement with our results.
\section{Conclusions}
In this work, we suggest that the homogeneous holographic models with broken translations and with global symmetries \color{black} might be described as \color{black} \textit{quasicrystals}. We show consistent evidence for this statement using the symmetry structure of the models and their hydrodynamic modes. In particular, we propose that the diffusive extra mode present in the longitudinal spectrum is a \textit{phason}, i.e. a diffusive Goldstone mode as shown in \cite{Donos:2019txg}, which is typical of \textit{quasicrystals}. Using this interpretation, and the methods of \cite{Hidaka:2019irz}, we are able to show directly the universal relation \eqref{toprove} between the phase relaxation rate, the Goldstone diffusion constant and the explicit breaking scale \cite{Amoretti:2018tzw}. Importantly, our interpretation implies that the longitudinal diffusive mode should be absent in  the periodic and non-homogeneous holographic lattices of \cite{Donos:2014yya,Donos:2015eew,Cremonini:2017usb,Cremonini:2016rbd,Andrade:2017leb,Krikun:2017cyw,Andrade2018}. This could serve as a definitive confirmation of our ideas \color{black}(in contrast to the proposal that such mode is related to the diffusion of vacancies)\color{black}.\\

In conclusions, in this work we provided a viable answer to the fundamental question of what all the homogeneous holographic models with broken translations are \color{black} and we showed how several existing puzzles could be solved using this interpretation. At this time, we cannot exclude the possibility of explaining these features using a different interpretation. Nevertheless, we are not aware of any able to pass all the non-trivial tests presented in this paper.\color{black} 
\section{Acknowledgments}
We thank D.Arean, Y.Hidaka, A.Jain, M.Landry, K.Landsteiner, Y.Minami, A.Zaccone and V.Ziogas for several discussions and correspondence on the topic. We thank M.Ammon, A.Donos, S.Grieninger, O.Pujolas and J.Zaanen for reading a preliminary version of this manuscript and providing useful comments. We thank Blaise Gouteraux and Wei-Jia Li for several interesting comments on the first version of this manuscript. The author acknowledges the support of the Spanish MINECO’s ``Centro de Excelencia Severo Ochoa'' Programme under grant SEV-2012-0249.

\bibliographystyle{apsrev4-1}
\bibliography{phason}

\end{document}